# Optical Bessel-like Beams with Engineered Axial Phase and Intensity Distribution

Sergej Orlov, Alfonsas Juršėnas, and Ernestas Nacius

*Center for Physical Sciences and Technology, Industrial laboratory for photonic technologies, Sauletekio av. 3, Vilnius, Lithuania*
*E-mail: Sergejus.orlovas@ftmc.lt*

Bessel beams are known for their property of maintaining propagation-invariant transverse intensity distribution. The true Bessel beams has unlimited energy, however the experimental realization of limited energy Bessel-like beams provides almost diffraction-free beams over a certain distance. These beams are applicable to microfabrication, particle manipulation etc. where long and narrow intense beam is needed. A type of laser beams that maintains narrow intensity peak for long distances are often called optical needles. In this paper we propose a construction of an "optical needle" using a superposition of vector Bessel beams and for flexible engineering of axial intensity profile. We investigate both theoretically and experimentally generation of various optical beams of different axial profiles, lengths, beam widths etc. We use a spatial light modulator as a toy model of an optical element to shape the spatial spectrum of the beam. Moreover, we demonstrate distortions introduced into the "optical needle" by a planar interface between two dielectric media: spherical aberration and astigmatism due to the nonzero inclination angle and discuss ways to correct them.
DOI: 10.2961/jlmn.2018.03.0017

## 1. Introduction

Laser beam shaping is an important technique used in modern laser beam applications such as light sheet microscopy, microfabrication and photopolimerisation to name a few. In situations where the same pattern is needed over long propagation distance it is advantageous to use non-diffracting beams. One of them is Bessel beam [1], which exhibits a long focal line with high length to width ratio [2]. Due to this property it can be perceived as an optical needle. Optical needle beams are advantageous to use in applications such as fabrication of long canals in bulk material, trapping many particles simultaneously [2-4].

In some cases the microfabrication process is sensitive to the polarization structure of the laser beam. Nonhomogeneous polarizations like azimuthal or radial have been shown to affect the efficiency of laser drilling procedure [5,6]. The polarization control of optical needle could also increase the speed of the laser microprocessing of the materials and will be discussed in this publication as an additional degree of freedom in the use of vectorial Bessel beams.

A multitude of different beam shaping methods can be used in order to experimentally obtain complex focal structures. One of the approaches is to use the spatial light modulator (which is a dynamic optical element but requires adjustments to the design in order to be flexibly used for high laser power applications), another example is geometrical phase elements (which are static but can sustain high laser power used in microfabrication applications), an example is given in [7].

In this work we present the formation of optical needle by superimposing many monochromatic vector Bessel beams to achieve desired axial intensity distribution. We analyze the generation of various optical needles with different axial intensity profiles. In order to test our theoretical results we compare our simulated beam intensity profiles with experimentally obtained ones. The experiment is conducted using spatial light modulator as beam spatial spectrum alternator. Furthermore we numerically demonstrate a method for compensating the axial intensity distortions caused by focusing the laser beam trough air-dielectric interface.

## 2. Control of an axial intensity of vector Bessel beams

In this chapter we introduce the vector description of Bessel beams [8] and a method of controlling the intensity distribution in the propagation direction of the beam [9,10].

### 2.1 Vector Bessel beams

Consider an ideal monochromatic scalar Bessel beam complex amplitude [1]

$$\psi_m(\rho,\phi,z) = J_m(k_\rho\rho)\exp(im\phi + ik_z z), \quad (1)$$

here $J_m$-Bessel function of order $m$, $\rho, \phi, z$ -circular cylinder spatial coordinates, $k_\rho, k_z$ - transverse and longitudinal of the wave vector **k**, $m$-topological charge. The scalar description can be extended to a full vector solution (denoted here as **M** and **N**) of Maxwell's equations in the free space using [8,11]

$$\mathbf{M} = \nabla \times (\mathbf{a}\psi_m), \qquad \mathbf{N} = \frac{1}{k}\nabla \times \mathbf{M}, \quad (2)$$

and vector solutions **M**, **N** are mutually orthogonal by the definition. The vector **a** enforces here vectorial boundary conditions that solutions **M**, **N** should obey [11]. Therefore the choice of **a** directly influences the polarization of the resulting electromagnetic field. For instance, with $\mathbf{a} = \mathbf{e}_x$ the field represented by the **M** solution is a vector Bessel beam, which is linearly polarized in the y direction

$$\mathbf{M} = \left[ ik_z \mathbf{e}_y J_0(k_\rho \rho) + \mathbf{e}_z \sin(\varphi) k_\rho J_1(k_\rho \rho) \right] e^{ik_z z} \qquad (3)$$

For $\mathbf{a} = \mathbf{e}_y$ the electromagnetic field represented by the **N** solution is also a linearly polarized in the y direction vector Bessel beam

$$\mathbf{N} = \frac{1}{k} \left\{ \mathbf{e}_x \frac{J_2(k_\rho \rho) k_\rho^2}{\rho^2} xy \right.$$
$$+ \mathbf{e}_y \left[ J_0(k_\rho \rho) k_z^2 + \frac{J_1(k_\rho \rho) k_\rho}{\rho} - \frac{J_2(k_\rho \rho) k_\rho^2}{\rho^2} x^2 \right], \quad (4)$$
$$\left. + ik_z \mathbf{e}_z \frac{J_1(k_\rho \rho) k_\rho y}{\rho} \right\} e^{ik_z z}$$

here $\mathbf{e}_x, \mathbf{e}_y, \mathbf{e}_z$ - Cartesian unit vectors.

An analysis of Eqs. (3,4) reveals that under tight focusing conditions, i.e. large numerical apertures, the longitudinal component of the electric field becomes larger.

### 2.2 Control of the axial intensity

We can express linearly polarized electromagnetic fields from Eqs. (3,4) close to the axis ($\rho \to 0$) as

$$\mathbf{M}(\rho \to 0) = ik_z \mathbf{e}_y e^{ik_z z}, \qquad (5)$$

$$\mathbf{N}(\rho \to 0) = \frac{k_\rho^2 + 2k_z^2}{2k} \mathbf{e}_y e^{ik_z z}. \qquad (6)$$

For the sake of brevity we will consider now a symmetrical mode, which is a sum of TM and TE modes, see Eqs. (5, 6). The axial intensity control of a linearly polarized Bessel beam superposition is achieved as follows. Let us assume a continuous representation of the axial beam profile via the spatial beam spectrum

$$A(K_z) = \frac{1}{2\pi} \int_{-\infty}^{+\infty} f(z) e^{-i K_z z} dz \qquad (7)$$

here $f(z)$ is a desired axial beam profile. We assume, that $k_z = k_{z0} + K_z$, where $k_{z0}$ is a carrier wave vector. If $k_{z0} \gg K_z$, we can rewrite (7) as a Fourier transform

$$\mathbf{\Psi}(\mathbf{r}) = \int_{-\infty}^{\infty} A(k_z) \psi(\mathbf{r}; k_z) dk_z, \qquad (8)$$

where $\psi(\mathbf{r}; k_z)$ is a symmetric combination of $\mathbf{M}(\mathbf{r}; k_z)$ and $\mathbf{N}(\mathbf{r}; k_z)$.

Thus, the monochromatic superposition of Bessel beams in Eq. (8) enables us to optically engineer the axial intensity distribution $f(z)$ of a Bessel-like optical beam. We note, that the case of a discreet Fourier transform for scalar beams with a Talbot effect was presented elsewhere [9,10].

### 3. Focusing through an air-dielectric interface

Focusing of the laser beam in the presence of an air-dielectric interface results in appearance of optical field distortions like spherical aberration, astigmatism and coma. Some of them are present only when the optical beam en-ters the second media perpendicularly to the interface, another do appear when the beam enters the media at an angle [12]. We note that usually Zernike polynomials are used to classify and characterize optical aberrations [13]. Here we aim for a more general approach, where we compensate all possible aberrations due to the dielectric interface without the exact classification of them using Zernike polynomials.

We assume that the focusing system is an aplanatic lens and use Ref. [12] as a guidance in our further derivations.

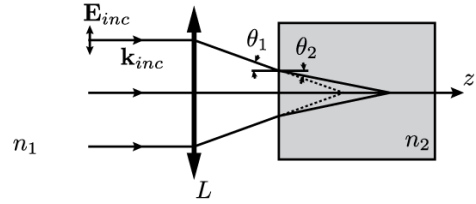

**Fig. 1** Focusing of beam $\mathbf{E}_{inc}$ trough air-dielectric interface using aplanatic lens $L$, the focusing angle in air is denoted $\theta_1$ and in the dielectric medium $\theta_2$ with refractive index $n_2$.

The focal field inside the dielectric material can be expressed as an integral over spatial angles of the plane waves [11]

$$\mathbf{E}_t = C_1 \int_0^{2\pi} \int_0^{\theta_{max}} \left\{ t^s(\theta_1)[\mathbf{V}(\theta_1,\phi)\hat{\phi}]\hat{\phi} + t^p(\theta_1)[\mathbf{V}(\theta_1,\phi)\hat{\theta}_1]\hat{\theta}_2 \right\} \times$$
$$\times e^{i(k_x x + k_y y) + i k_{z2} z} \sin(\theta_2)\sqrt{\cos(\theta_2)} d\theta_2 d\phi, \qquad (9)$$

$t^s$, $t^p$ are Fresnel coefficients [12], $C_1 = -ikfe^{-ikf}/2\pi$ is a constant, $f$ - focal length of an aplanatic lens, $\theta_1$, $\theta_2$, $\phi$ angles of the plane wave components, see Fig. 1, $\hat{\theta}_1$, $\hat{\theta}_2$, $\hat{\phi}$ - spherical unit vectors, **V** - denotes the spatial spectra of the vector beam – a linear combination of $\mathbf{N}(\theta_1,\phi)$ and $\mathbf{M}(\theta_1,\phi)$ - angular spectra of the solutions from Eq. (3,4).

An analysis of the expression (9) reveals that main causes for the distortions of the initial field after entering the second medium is due to Fresnel coefficients and due to the the Snell's law, which changes the angular dispersion of the optical beam.

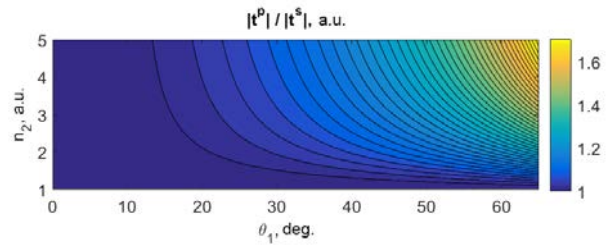

**Fig. 2** Dependency of the ratio $|t^p|/|t^s|$ on angles $\theta_1$ and refractive index $n_2$ of the dielectric medium.

We plot now the ratio $|t^p|/|t^s|$ of the Fresnel coefficients for s- and p- polarization as a function of angles and refractive indices of the second medium, see Fig. 2. We can observe a rather large region in the parameter space $\theta_1, n_2$, where the polarization structure of the spatial spectra won't be significantly affected by the interface itself be-cause $|t^p|/|t^s| \approx 1$. In this region the distortion in the engineered profile will be caused mainly by the Snell's law. They can be accounted for by adjusting the angular dispersion of the entering optical beam using the interchange of variables $\theta_1 \mapsto \theta_2$ and $\mathbf{V}(\phi,\theta_1) \mapsto \mathbf{V}(\phi,\theta_2)$, see Eq. (9). After aforementioned substitution we obtain the following expression

$$\mathbf{E}_t = C_1 \int_0^{2\pi}\int_0^{\theta_{max}} \left\{ t^s(\theta_1)[\mathbf{V}(\theta_2,\phi)\hat{\phi}]\hat{\phi} + t^p(\theta_1)[\mathbf{V}(\theta_2,\phi)\hat{\theta}_2]\hat{\theta}_2 \right\} \times$$
$$\times e^{i(k_x x + k_y y) + ik_{z2} z} \sin(\theta_2)\sqrt{\cos(\theta_2)}d\theta_2 d\phi. \quad (10)$$

Expressions in Eqs. (9, 10) can be numerically evaluated, so we choose an axial distribution described by a function $f(z) = \Pi(Lz - L_1) + \Pi(Lz - L_2)$, where $\Pi(z)$ is a boxcar (or a rectangular) function, $L$ is its length, $L_1$ and $L_2$ are positions. The intensity profiles in the longitudinal plane $(x,z)$ along with the intensity profiles of a spatial spectra are depicted in the Fig. 3.

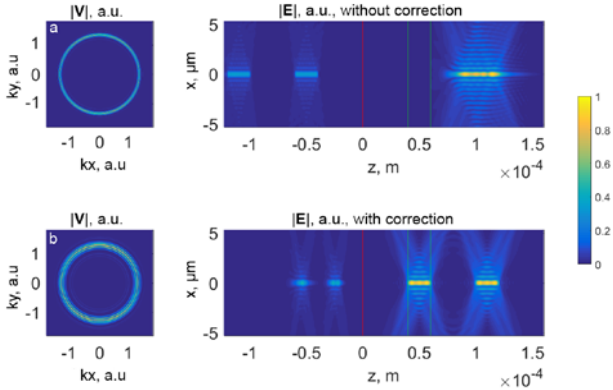

**Fig. 3** Intensity distribution of an optical needle with a two-step axial profile in the longitudinal plane (on the right) for two cases: first line – direct focusing onto the interface, second line – direct focusing but with corrections applied. Intensity distribution in the Fourier plane for both cases (on the left). The refractive index $n_2 = 1.8$, the position of the interface is denoted by a red line ($z=0$), expected length $L$ of the optical needle is denoted by green vertical lines. The maximal angle in the spatial spectra is $\theta_{max} = 45°$.

On the left side of the interface ($z<0$) (see Fig. 3) a reflected beam is observed in both cases. We note, that in the case of direct focusing onto the interface, the first rectangular profile is enlarged, shifted into the material and its shape is somehow distorted. However, when the spatial spectra is modified to account for the presence of the interface, the expected axial profile is restored.

### 3.1 Focusing into medium under an angle

Let us consider a case when the angle $\alpha$ between the optical axis of the beam and the normal to the interface is nonzero. Expressions for the electric field in this case can be found from the Eq. (10) by applying the rotation transformation to the incident field and modifying the spatial spectra, so the $k$-vectors are properly transformed. More details on this procedure is out of the scope of this report, but details can be found in Ref. [12].

In this case, we use a slightly another axial intensity profile $f(z) = \Pi(Lz - L_1)$ and present a comparison of two cases with different angles between the optical axis and the interface, see Fig. 4. We observe that the spatial spectra can be readjusted in order to ensure the expected length of the optical needle bot for the case on zero incidence angle and for non-zero.

Nevertheless, we do observe that the transverse distribution (computed along the red dashed line) is asymmetrically distorted, when focusing into the medium with non-zero incidence angle.

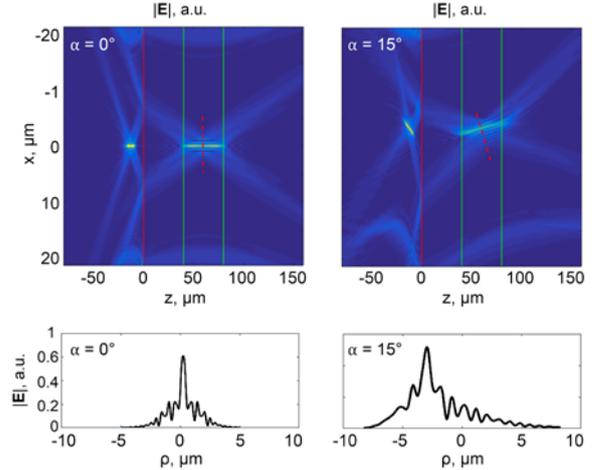

**Fig. 4** Intensity distribution of an optical needle with a step-like axial profile in the longitudinal plane (first row) for two cases: direct focusing but with corrections applied for angles $\alpha = 0°, 15°$. Second row, intensity distributions on the red dotted line for two cases depicted above. The position of the interface is denoted by a red line ($z=0$), expected length $L$ of the optical needle is denoted by green vertical lines. The maximal angle in the spatial spectra is $\theta_{max} = 45°$. The refractive index of the second medium $n_2 = 4$.

### 4. Experimental results
#### 4.1 Setup of the experiment

Numerical simulations were verified experimentally using an optical setup depicted in Fig. 5, where the spatial light modulator (SLM) was used to alter the spatial spectrum of the beam. The procedure is as follows:

1. A linearly polarized and expanded light beam (at the wavelength of 532 nm) reaches the matrix of the SLM at angle of incidence of zero degrees.

2. It is reflected from the liquid crystal mask of the SLM, where the phase mask encodes both amplitude and phase spatial distribution. The 4f lens system (lenses L2 and L3) is used to transfer the spectral image from the SLM to a lens L4.

3. The lens L4 makes a Fourier transform of the beam generated on the SLM. The size of the liquid crystal mask of the SLM ($x_s$, $y_s$) and focal length of the Fourier lens $f_4$ determine spatial frequencies $f_x$ and $f_y$, which can be achieved in the setup.

4. Knowing the window of spatial frequencies ($f_x$, $f_y$), one can relate them to the maximal values of transverse wave vectors $k_x$, $k_y$ via $k = 2\pi f$. The pixel size $dx \times dy$ of the SLM determines the pixel size $dk_x \times dk_y$ of the spatial spectrum picture, which we obtain from Eqs. (7,8).
5. The complex valued spatial spectrum has to be encoded to a phase-only picture, suitable for our SLM. The amplitude modulation is implemented here by using a checkerboard mask method, where groups of four phase-only pixels emulate a single pixel with arbitrarily chosen amplitude and phase [14]. In this method the amplitude A of the complex amplitude $A\exp(i\varphi)$ is encoded as the sum of two different phases for a 4x4 checkerboard: $\varphi_1 = \varphi - \arccos A$ and $\varphi_2 = \varphi + \arccos A$.
6. Thus, a desired beam profile is generated in the focal region of the lens L4. It is captured and recorded afterwards by the imaging system. A moving linear translation stage with mounted imaging system is moved along z axis.
7. While the stage moves, the transverse intensity data at different distances from focal plane is imaged onto the CCD matrix of the camera, this image is recorded and processed after-wards by a personal computer.

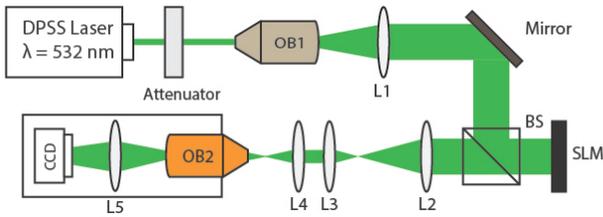

**Fig. 5** Experimental setup: CW laser, lenses (L1-L5), objectives (OB1 and OB2), spatial light modulator (PLUTOVIS-006-A, HOLOEYE Photonics AG) and CCD camera.

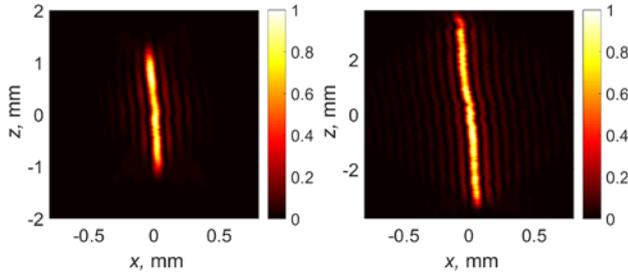

**Fig. 6** Intensity distributions of two experimentally observed optical needles with different lengths: $L_1 = -0.2$ mm and $L_2 = 0.2$ mm (left) and $L_1 = -0.6$ mm and $L_2 = 0.6$ mm (right). Note that due to magnification $z$ scale is 6 times bigger.

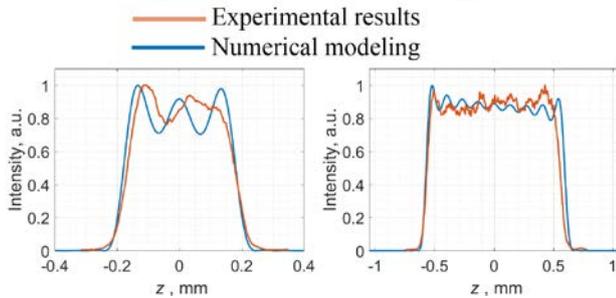

**Fig. 7** Comparison of numerically calculated and experimentally observed axial intensity profiles of beams: $L_1 = -0.2$ mm and $L_2 = 0.2$ mm (left) and $L_1 = -0.6$ mm and $L_2 = 0.6$ mm (right).

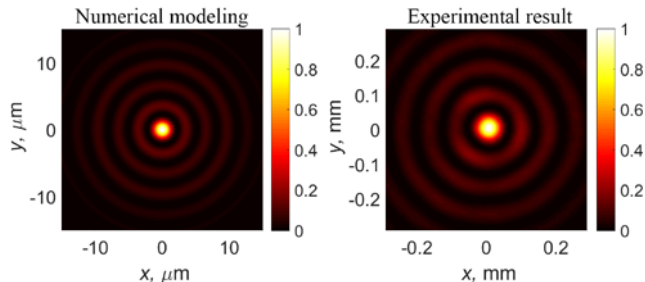

**Fig. 8** Intensity distribution of numerically simulated (left) and experimentally observed (right) optical needles in the transverse planes in the plane $z = 0$ mm, $L_1 = -0.2$ mm and $L_2 = 0.2$ mm.

Our optical system achieves transverse magnification of approximately x27.4 and a longitudinal magnification of approximately x6.

### 4.2 Experiments with controllable axial intensity

Optical needles with various potentially interesting axial intensity profiles can be easily formed using the method described in section 2.2. We start by a rectangular function (boxcar) $f(z) = \Pi(Lz)$ with different lengths $L$. Such needles can have different axial profiles, lengths, beawidths etc. The start of the rectangular function is at $L_1 = -L/2$ and the end is at the $L_2 = L/2$, so that $L = L_2 - L_1$, and we can control beam length in the $z$ direction. An example of two needles with different lengths are depicted in the Fig. 6. The figure 7 presents comparison of the maximum axial intensity distribution over the propagation distance between the numerical simulation (blue curve) and experiment (red curve).

Although, the intensity modulation peaks observed experimentally not always actually match intensity modulation in the numerically modelled curves, which could be attributed to the use of the discretized SLM grid.

We observe a very similar intensity growth and decay locations and speed. In overall, a relatively good agreement between our numerical expectations and experimental results is observed.

Both axial profiles closely match numerical curves for the axial intensity distribution, see Fig. 7. A comparison of the transverse intensity distributions at the same plane $z = 0$ mm is presented in the Fig. 8. Once again, it is clearly seen that the degree of conformity between numerical modelling and experimental results is quite high.

### 4.3 Experiments with more complex axial profiles

Lastly, we analyze more complex axial profiles. For the start we analyze creation of the profile, which was already presented in the section 3– the two step-like function

$$f(z) = \Pi\left(L_2 z - L_1 z - \frac{L_1 + L_2}{2}\right) + \Pi\left(L_4 z - L_3 z - \frac{L_3 + L_4}{2}\right). \quad (11)$$

This particular profile consists of two different length boxcars. We have also analyzed more elaborate axial intensity profile containing two exponentials, one rising and another falling:

$$f(z) = \exp[(z - L_2)]\Pi\left(L_2 z - L_1 z - \frac{L_1 + L_2}{2}\right)$$
$$+ \exp[-(z - L_3)]\Pi\left(L_4 z - L_3 z - \frac{L_3 + L_4}{2}\right). \quad (12)$$

Experiment and numerical simulation results depicting the axial intensity are presented in the figure 9. Our numerical expectations are compared with our experimental results in the same Figure. Here, numerical expectations are show in the blue, and our experimental measurements are depicted in red, see Fig. 9.

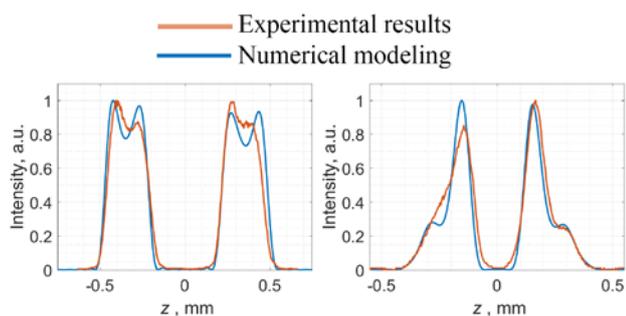

**Fig. 9** Intensity distribution of experimentally obtained and numerical calculated optical needles on the z axis. Parameters are $L_1 = -0.5$ mm, $L_2 = -0.2$ mm, $L_3 = 0.2$ mm, $L_4 = 0.5$ for the two step function (left) and $L_1 = -0.5$ mm, $L_2 = -0.1$ mm, $L_3 = 0.1$ mm, $L_4 = 0.5$ (right) for two adjacent exponential profiles.

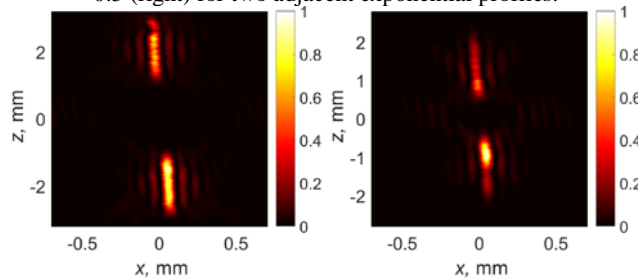

**Fig. 10** Two-dimensional intensity distribution of the structures represented in the Fig. 11.

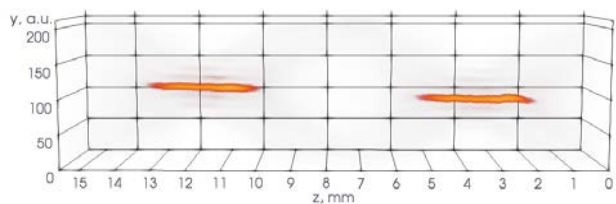

**Fig. 11** Three-dimensional view of intensity distribution depicted in the Fig. 11 (left picture).

We note, that we have observed interference between two adjacent optical needles, which distorts the axial intensity profile near the focal plane ($z=0.0$ mm).

Our preliminary analysis reveals, that this is happening due to the physical limitations of our experimental system – the dimensions of the liquid crystal matrix in the SLM are too big to properly represent the fine phase modulation in the spatial spectrum. A two dimensional view on the intensity distributions can be found in the Figure 10.

Lastly, a three dimensional representation of the intensity isosurfaces is depicted for this complex structure in Fig. 11.

Using this experimental setup we were able to control not only to the individual on-axis intensity distribution of an optical needle, but also the individual axial position of each individual optical needle and their relative amplitudes in a superposition of many individual beams.

## 5.  Conclusions and discussions

In conclusion, we have presented a flexible technique, which enables us to create experimentally controlled axial intensity distributions of linearly polarized optical needles. We have briefly discussed a creation of complex axial structures, which contains a number of individual optical beams, with independent axial intensity profiles, axial positions and amplitudes.

Moreover, we have also discussed in great detail theoretical methods, which would let us compensate the spherical aberration, some astigmatism related effects due to the focusing the beam to the sample with non-zero angles of incidence. Those effects become more important nowadays in such applications like manufacturing high-aspect-ratio through-silicon vias, see Ref. [15].

These theoretical considerations will be also important in the near future while adapting techniques presented in this short report for the actual laser microprocessing of transparent materials.

Further results on this approach together with an experimental implementation of the concept using geometrical phase elements (produced by "Altechna R&D" using birefringent nanogratings inscribed in the glass volume, see Refs. [7,16] for more details) will be presented elsewhere.

## Acknowledgements

This research is funded by the European Social Fund according to the activity 'Improvement of researchers' qualification by implementing world-class R&D projects' of Measure No. 09.3.3-LMT-K-712